# AN EFFICIENT CNTFET-BASED 7-INPUT MINORITY GATE

Samira Shirinabadi Farahani[1,2], Ronak Zarhoun[1,2],
Mohammad Hossein Moaiyeri[1,2] and Keivan Navi[1,2]

[1]Faculty of Electrical and Computer Engineering,
Shahid Beheshti University, G.C., Tehran, Iran
[2]Nanotechnology and Quantum Computing Lab,
Shahid Beheshti University, G.C., Tehran, Iran
navi@sbu.ac.ir

## ABSTRACT

*Complementary metal oxide semiconductor technology (CMOS) has been faced critical challenges in nano-scale regime. CNTFET (Carbon Nanotube Field effect transistor) technology is a promising alternative for CMOS technology. In this paper, we proposed a novel 7-input minority gate in CNTFET technology that has only 9 CNTFETs. Minority function is utilized in the voting systems for decision making and also it is used in data mining. This proposed 7-input minority gate is utilized less fewer transistors than the conventional CMOS method which utilizes many transistors for implementing sum of products. By means of this proposed 7-input minority gate, a 4-input NAND gate can be implemented, which gets better the conventional design in terms of delay and energy efficiency and has much more deriving power at its output.*

## KEYWORDS

*Nanoelectronics, Minority function, CNTFET technology, Logic gates.*

## 1. INTRODUCTION

As Gordon Moore predicted in 1965, the number of transistors in chips duplicates every 18 months [1]. Therefore, for embedding more transistors on chips, their feature size should be reduced. Complementary metal oxide semiconductor (CMOS) is no more suitable for near future nano-scale regime. By Scaling down the feature size of CMOS technology, many challenges appear. Lithography limitations, large parametric variations, high power density are some of these critical challenges. To overcome these difficulties and challenges in sub-nanometre scale new technologies have emerged. Among these novel technologies such as, carbon nanotube field effect transistor (CNTFET), quantum-dot cellular automata (QCA) Benzene ring transistors and single electron transistor (SET), CNTFET looks to be more feasible because of its CMOS-like structure [2-7].

Minority function which has complementary behaviour of majority function, is a complete function gate because it has capability of implementing other gates such as NAND and NOR. A 7-input minority gate has 7 binary inputs and its single output will be equal to '1' when 3, 2, 1 or none of the inputs are '0' and if not the output will be '0' [8-10]. Minority vote of inputs can be considered as complementary of carry-out of arithmetic summation [11]. Minority function is used for data mining and it is also used in the voting systems [8,12]. Design of a 7-input minority function seems to be important because in some cases fast and minimum cost logical gates with large fan-in are required. By means of designing 7-input minority function gate deriving 4-input logical gates such as 4-input AND, OR, NAND and NOR becomes easier and more reliable since majority-based structures are utilized not only in conventional fault-tolerant architectures but also in new nano-scale technologies [4]. Implementing 7-input minority function in conventional method by using sum of products (SOP) is difficult and costly, especially when it is implemented in conventional CMOS style that the number of transistors multiplied with two due to the pull-down and pull-up networks.

The number of SOPs and the number of transistors can be calculated as describe in Eqn. 1 and 2, respectively.

$$S = \#SOPs = \sum_{i=\lceil\frac{n}{2}\rceil}^{n} \binom{n}{i} = \sum_{i=\lceil\frac{n}{2}\rceil}^{n} \frac{n!}{i!(n-i)!} \qquad (1)$$

Where n is the number of inputs.

$$T = \# \text{ Transistors} = 2 \cdot S \cdot n \qquad (2)$$

In this work, a new 7-input minority function gate is proposed, which is just composed of high-speed CNTFETs.

The reminder of this paper is organized as follows: in section 2, a brief review of CNTFET technology is provided and the novel 7-input minority function gate is proposed in section 3. Simulation results and the conclusion are presented in section 4 and section 5 respectively.

## 2. CNTFET TECHNOLOGY

Carbon nanotube (CNT) consists of a graphene sheet that rolled into a cylindrical structure. CNTs can be classified into two groups, SWCNT and MWCNT. The first one has only one cylinder and the former has multi cylinders [13]. Each SWCNT has a two dimensional vector $(\overrightarrow{n_1}, \overrightarrow{n_2}) = (n_1\overrightarrow{a_1}, n_2\overrightarrow{a_2})$, called chiral vector that specifies its electrical properties [14]. The SWCNT has the zigzag structure when $n_1=0$ or $n_2=0$, and if $n_1=n_2$, the SWCNT has the armchair structure. Chiral vector defines whether a SWCNT is semiconducting or not. If $|n_1 - n_2| = 3k (k \in Z)$ the SWCNT is metallic and conducting, otherwise it is semiconducting [14]. For determining the diameter of a SWCNT, we can draw a carbon molecule as a regular hexagon in a circle as shown in Fig. 1.

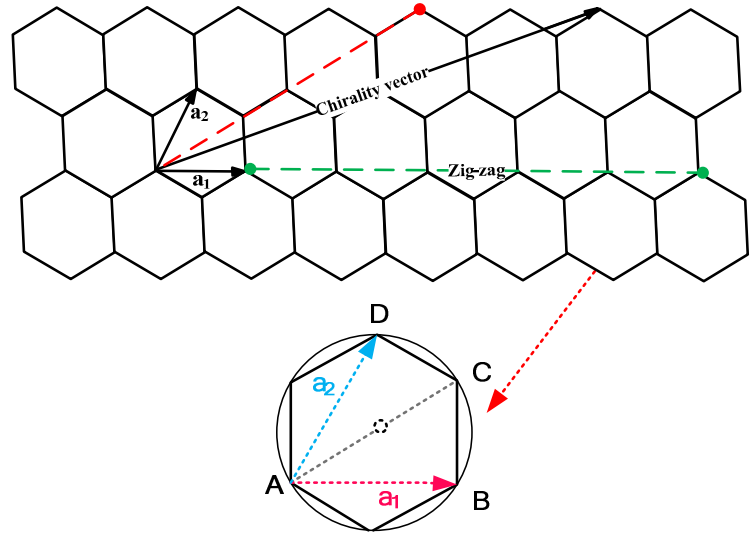

Figure 1. Unrolled graphite sheet

In this figure triangular ABD is isosceles, so $a_1 = a_2$. By considering the rectangular triangle ABC and by means of triangular relationship $|a_1|$ and $|a_2|$ is determined as:

$$a_1 = a_2 = 2a_0 \sin(60°) = \sqrt{3}a_0 \qquad (3)$$

The chiral vector is calculated as:

$$Ch^2 = a_1^2 n_1^2 + a_2^2 n_2^2 + 2a_1 a_2 n_1 n_2 \cos(60°) \qquad (4)$$

$$Ch = \sqrt{3}a_0 \sqrt{n_1^2 + n_2^2 + n_1 n_2} \qquad (5)$$

Where, $a_0$ shows the carbon to carbon atom distance. The diameter of a CNT can be calculated according to the following relation [15]:

$$D_{CNT} = \frac{\sqrt{3}a_0 \sqrt{n_1^2 + n_2^2 + n_1 n_2}}{\pi} \qquad (6)$$

Structure and operation of the carbon nanotube field effect transistors are more similar to traditional silicon transistors but the conduction channel in CNTFETs is consists of semiconducting SWCNTs [5,6]. Threshold voltage of a CNTFET is approximately calculated as [15]:

$$Vth \approx \frac{E_g}{2e} = \frac{\sqrt{3}}{3} \frac{a_0 V_\pi}{eD_{CNT}} \approx \frac{0.43}{D_{CNT}(nm)} \qquad (7)$$

Structure and operation of the carbon nanotube field effect transistors are more similar to traditional silicon transistors but the conduction channel in CNTFETs is consists of semiconducting SWCNTs [5,6]. Threshold voltage of a CNTFET is approximately calculated as [15]:

$$\text{Vth} \approx \frac{E_g}{2e} = \frac{\sqrt{3}}{3} \frac{a_0 V_\pi}{e D_{CNT}} \approx \frac{0.43}{D_{CNT}(\text{nm})} \qquad (7)$$

Where $V_\pi$ is the carbon π-π bond energy amount in the tight bonding model and is equal to 3.033eV and e is the unit charge of electron.

Many effective efforts have been made for fabricating CNFET based integrated circuits. For instance, in [16] fabrication of CMOS-style CNFET-based inverter has been reported and in [17] fabrication of VLSI-compatible CNFET-based commonly used combinational and sequential circuits has been reported.

## 3. PROPOSED WORK

The proposed 7-input minority function gate, shown in Fig. 2, has seven CNTFETs that each of them acts as a capacitor [18] and are connected to a CNTFET-based inverter. The truth table of 7-input minority function has $2^7$=128 rows. For reducing the size of the truth table, the summation of inputs, as demonstrated in Table 1, is used.

Table 1. Truth Table of 7-input minority function

| $\sum$ in | 7-minority |
|-----------|------------|
| 0 | 1 |
| 1 | 1 |
| 2 | 1 |
| 3 | 1 |
| 4 | 0 |
| 5 | 0 |
| 6 | 0 |
| 7 | 0 |

Eqn. 8 formulates the functionality of presented truth table.

$$\text{out} = \begin{cases} 1 & \sum \text{in} \leq \left\lfloor \dfrac{\#\text{inputs}}{2} \right\rfloor \\ 0 & \text{otherwise} \end{cases}$$
$$= \begin{cases} 1 & \sum \text{in} \leq 3 \\ 0 & \text{otherwise} \end{cases} \qquad (8)$$

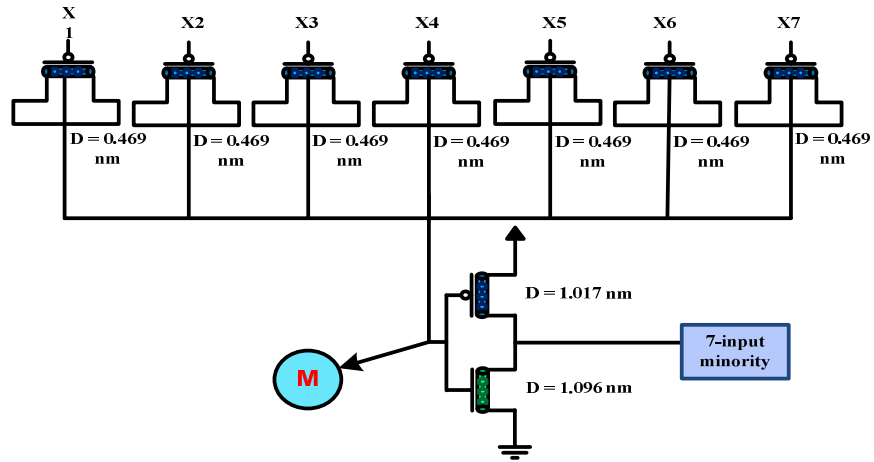

Figure 2.  Proposed design of 7-input minority function

Now we describe the structure of our proposed circuit. The middle point of the circuit, shown by 'M' in Fig. 2, has a voltage corresponding to the scaled sum of inputs. As a result, the voltage level of this point can be expressed by Eqn. 9.

$$V_m = \frac{1}{\#\text{inputs}} \sum_{i=1}^{7} x_i \qquad (9)$$

A minority function is inherently a voter, but instead of reflection of the majority vote, it shows the minor vote at its output, therefore an inverter with a specified threshold is utilized to generate the expected output:

$$\text{Inverter threshold} \propto \left\lfloor \frac{\#\text{inputs}}{2} \right\rfloor \qquad (10)$$

The voltage transfer characteristics (VTC) curve of the inverter is shown in Fig. 3.

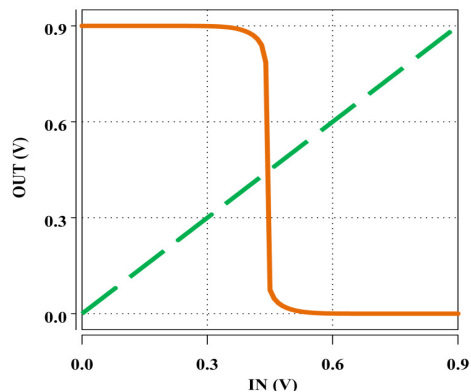

Figure 3.  VCT of the inverter

The 7- input minority gate can be used as the building block of efficient 4-input NAND gate and 4-input NOR gate. The way of designing these basic 4-input logic gates based on the proposed 7-input minority gate is presented in Fig. 4.

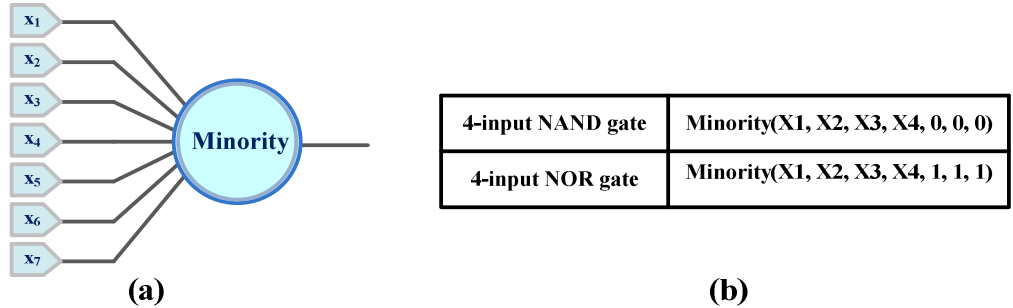

Figure 4. Schematic of a 7-input minority gate (a). NAND and NOR gates based-on minority (b).

## 4. SIMULATION RESULTS

The proposed 7-input minority function gate are simulated with synopsis Hspice 2008 with the 32 nm CMOS technology and Compact SPICE model for 32 nm CNTFETs ($L_g$ = 32 nm) at room temperature at 0.9 V supply voltage and with a 2fF output load capacitance [19, 20]. This model was designed for unipolar, MOSFET-like CNTFET devices and each transistor can have one or more CNTs. Schottky barrier effect, parasitic including gate and source/drain resistance and capacitance and CNT charge screening effect are considered in this model. In Table 2, the parameters of this model with their values and a brief explanation for each parameter are provided [21].

The simulation results, including the propagation delay, the average power consumption and the average energy consumption are calculated and presented Table 3.

Table 2. CNTFET model parameter.

| Parameter | Explanation | Value |
|---|---|---|
| $L_{ch}$ | Physical channel length | 32 nm |
| $L_{geff}$ | The mean free path in the intrinsic CNT channel | 100 nm |
| $L_{ss}$ | The length of doped CNT source-side extension region | 32 nm |
| $L_{dd}$ | The length of doped CNT drain-side extension | 32 nm |
| $K_{gate}$ | The dielectric constant of high-k top gate dielectric material | 16 |
| $T_{ox}$ | The thickness of high-k top gate dielectric material | 4 nm |
| $C_{sub}$ | The coupling capacitance between the channel region and the substrate | 40 pF/m |
| Efi | The Fermi level of the doped S/D tube | 6 eV |

Table 3. Simulation results of the proposed minority gate versus frequency variations.

| Frequency (MHz) | Delay ($\times 10^{-12}$ s) | Power ($\times 10^{-6}$ W) | Energy Consumption ($\times 10^{-17}$ J) |
|---|---|---|---|
| 250 | 23.0 | 2.25 | 5.18 |
| 500 | 22.2 | 3.81 | 8.45 |
| 1000 | 23.4 | 3.45 | 8.07 |

The energy consumption is the product of the average power and the propagation delay of the circuits and makes a trade off between these two major performance metrics. Simulation results for different amount of VDD are presented in Table 4.

Table 4. Simulation results of the proposed minority gate versus power supply variations.

| VDD (V) | Delay ($\times 10^{-12}$ s) | Power ($\times 10^{-6}$ W) | Energy Consumption ($\times 10^{-17}$ J) |
|---|---|---|---|
| 0.8 | 58.3 | 0.792 | 4.61 |
| 0.9 | 23.0 | 2.25 | 5.18 |
| 1 | 14.9 | 6.96 | 10.3 |

Moreover, as an instance, the performance of a 4-input conventional CMOS NAND gate, shown in Fig. 5(a), is compared with a 4-input NAND gate derived from the proposed 7-input minority function gate, shown in Fig. 5(b). According to Fig. 4(b) and as demonstrated in Fig. 5(b), three inputs of the minority gate should be connected to ground. For implementing with CNTFETs, a CNTFET, whose number of tubes is almost 3 times greater than CNTFETs used for four inputs, can be utilized [21].

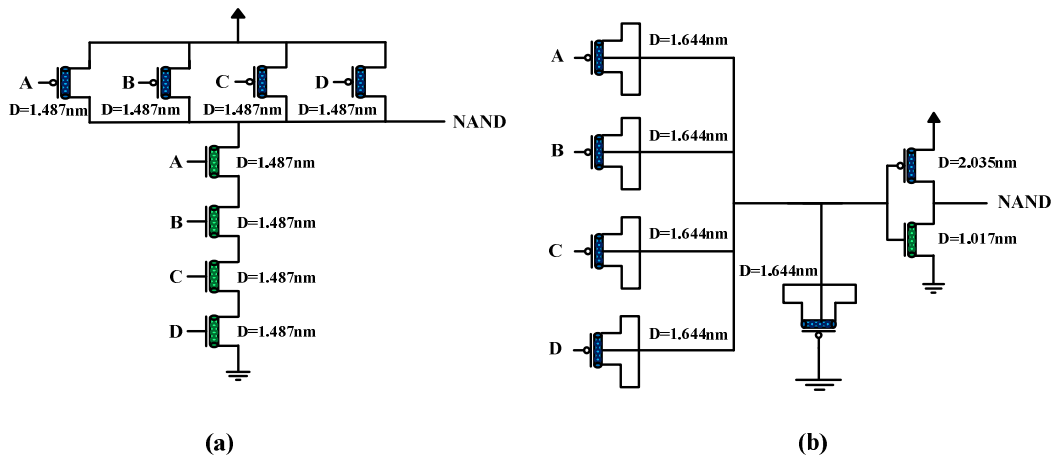

**(a)**                                                                    **(b)**

Figure 5. Conventional 4-input NAND (a). Proposed minority-based 4-input NAND gate (b)

Simulation results, listed in Table 5, demonstrate 22.67% improvement in terms of energy consumption and 57.67% in terms of delay using a large load capacitance (20 fF) at 0.9 V, which demonstrates the high driving capability of the proposed design. The propagation delay and energy consumption of the designs are plotted versus load capacitance in Fig. 6 and Fig. 7, respectively.

Table 5. Simulation results of the minority-based and conventional 4-input NANDs

| VDD (V) | Delay ($\times 10^{-12}$ s) | | Energy consumption ($\times 10^{-17}$ J) | |
|---|---|---|---|---|
| | Conventional NAND | Proposed NAND | Conventional NAND | Proposed NAND |
| 0.8 | 53.5 | 25.8 | 10.3 | 6.83 |
| 0.9 | 49.7 | 21.1 | 12.1 | 9.29 |
| 1 | 46.4 | 18.8 | 14 | 12.5 |

## 5. CONCLUSION

In this work, a fully CNTFET-based 7-input minority function gate is proposed, which can be also used for designing efficient 4-input logic gates. The proposed 7-input minority function gate, has only 9 CNTFET, while for implementing this function in the conventional way, more CNTFETs ( $2 \cdot S \cdot n = 896$ ) are required. Therefore this novel 7-input minority gate shows more than 98% improvement in terms of number of utilized CNTFETs. The proposed gate is logically complete and the other logic gates can be derived using this gate. As it is shown by the simulation results, as an instance the 4-input NAND gate based on the proposed design outperforms the conventional design in terms of delay and energy efficiency and has much more deriving power at its output.

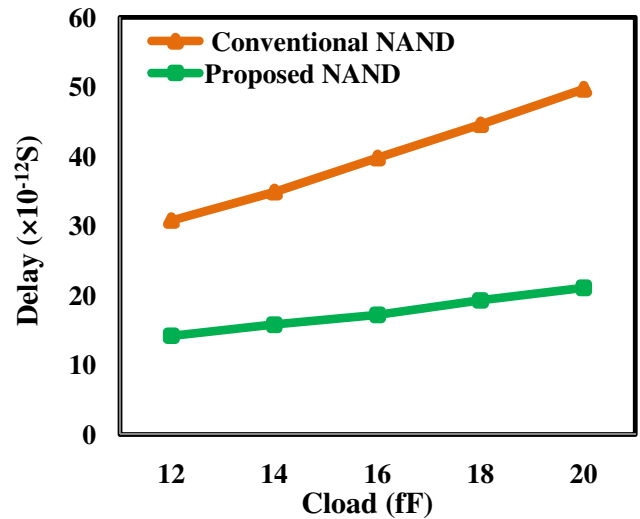

Figure 6. Delay versus load capacitance.

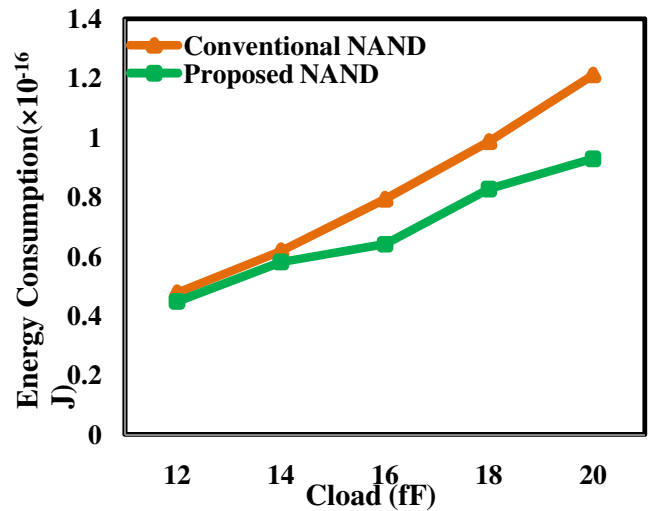

Figure 7. Energy consumption versus load capacitance.